\documentclass{aa}  

\usepackage{graphicx}
\usepackage{txfonts}
\usepackage{subcaption}         % necessary for continued figures, example in section 3
\usepackage{lscape}             % to rotate a single page table, example in appendix.
\usepackage{placeins}           % useful with \FloatBarrier, to keep 
\usepackage{orcidlink}
\usepackage{bm}
\usepackage{empheq}

\newcommand\rd{{\rm d}}

\newcommand\btheta{\bm{\theta}}

\newcommand\bx{\bm{x}}

\newcommand\ri{{\rm i}}

\newcommand\bA{\bm{\mathcal{A}}}

\newcommand\vect[1]{\boldsymbol{#1}}

\newcommand{\uvect}[1]{\hat{\vect{#1}}}

\newcommand\e[1]{_{\mathrm{#1}}}
\newcommand\h[1]{^{\mathrm{#1}}}

\newcommand{\delimiters}[4][]{
\ifthenelse{ \equal{#1}{1} }{  #2 #3 #4  }
					{ \ifthenelse{\equal{#1}{2}}{ \big#2 #3 \big#4 }
						{ \ifthenelse{\equal{#1}{3}}{ \Big#2 #3 \Big#4 }
							{ \ifthenelse{\equal{#1}{4}}{ \bigg#2 #3 \bigg#4 }
								{ \ifthenelse{\equal{#1}{5}}{ \Bigg#2 #3 \Bigg#4 }
									{ \left#2 #3 \right#4 }
								}
							}
						}
					}
													}
													
\newcommand{\pa}[2][]{\delimiters[#1]{(}{#2}{)}}
\newcommand{\pac}[2][]{\delimiters[#1]{[}{#2}{]}}

\usepackage{hyperref}
\usepackage{cleveref}
\begin{document}

%%%%%%%%%%%%%%%%%%%%%%%%%%%%%%%%%%%%%%%%
% if you use custom commands in your title,
% ensure to check your title when submitting!
%%%%%%%%%%%%%%%%%%%%%%%%%%%%%%%%%%%%%%%%

   \title{Degeneracies and modelling choices in double-plane time-delay cosmography}

%%%%%%%%%%%%%%%%%%%%%%%%%%%%%%%%%%%%%%%%
% Please separate each author with the \and command
%
% Please do not include ORCIDs next to author names.
% Only ORCIDs authenticated by individual authors in EDPS
% editorial system will be taken into account.
% ORCIDs included here will be removed.
%%%%%%%%%%%%%%%%%%%%%%%%%%%%%%%%%%%%%%%%

   \author{Daniel Johnson \inst{1} \fnmsep\thanks{Corresponding author: daniel.johnson@umontpellier.fr} \orcidlink{0000-0002-0311-2513}
        \and Pierre Fleury \inst{1} \orcidlink{0000-0001-9292-3651}
        \and Martin Millon \inst{2,3} \orcidlink{0000-0001-7051-497X}}

   \institute{Laboratoire Univers et Particules de Montpellier (LUPM), 
CNRS \& Université de Montpellier (UMR-5299),
Parvis Alexander Grothendieck, F-34095 Montpellier Cedex 05, France
   \and D\'epartement de Physique Th\'eorique, Universit\'e de Gen\`eve, 24 quai Ernest-Ansermet, CH-1211 Gen\`eve 4, Switzerland
\and Institute for Particle Physics and Astrophysics, ETH Zurich,
Wolfgang-Pauli-Strasse 27, CH-8093 Zurich, Switzerland}

   \date{Received September 30, 20XX}

% \abstract{}{}{}{}{}
% 5 {} token are mandatory
 
  \abstract
   {Double-plane gravitational lensing is a rare but increasingly observed phenomenon in which the light from a distant source is lensed by two foreground objects at different redshifts. Such systems can be used to provide simultaneous constraints on the Hubble constant~$H_0$ and the dark-energy equation of state, independent of and complementary to other probes. However, just as for single-plane gravitational lenses, the precision of these constraints is limited by the so-called mass-sheet degeneracy (MSD) -- a fundamental limit to the knowledge of the mass profiles of lens galaxies and the line of sight that can be obtained from imaging constraints alone. In this work, we show explicitly how contributions from the line of sight appear in double-plane systems. Because these contributions modify angular diameter distances, we argue that cosmological priors should not be used to simply fix the ``cosmological scaling factor'', a ratio of angular diameter distances which is key to the modelling of double-plane lenses. Motivated by this fact, we generalise the double-plane MSD to account for this uncertainty in the scaling factor. While this complicates the time-delay function, we show that, using the ``unfolding relation'', a geometric relation between distances which holds even in the presence of line-of-sight corrections, the uncertainty in the Hubble constant reduces to the familiar mass-sheet transformation of the first lens plane, and a line-of-sight contribution between the observer and the second lens plane. Our main message is therefore a prescription for reducing the degrees of freedom within double-plane models, while still safely accounting for the MSD in measurements of $H_0$.}


   \maketitle
   \nolinenumbers

%%%%%%%%%%%%%%%%%%%%%%%%%%%%
\section{Introduction}
\label{sec:introduction}
%%%%%%%%%%%%%%%%%%%%%%%%%%%%
%
The path taken by a light beam propagating through the universe is determined by the spacetime geometry along its path, and thus image positions and arrival times carry information about both the background cosmology and the matter distribution through which the beam has passed. Lens modelling is the attempt to tease out this information, and determine parameters of the source, background cosmology, and lens mass which would conspire to reproduce the images we see in our telescopes. 

Two of the most competitive cosmological applications of strong gravitational lensing are found in the fields of time-delay cosmography and double-plane lensing. The arrival time of a light ray depends on the lens mass distribution and on the angular diameter distances from and to the lens and source, which in turn depend on the cosmic expansion rate when expressed in terms of their observed redshifts. If a multiply-lensed source is sufficiently bright, compact and variable, such that changes in its flux are observable on reasonable timescales, and if the lensing potential can be well-constrained, the relative delays between the multiple images enable a measurement of the Hubble constant $H_0$, which is completely independent of distance-ladder or model-dependent early universe measurements \citep{Refsdal_1964,Treu_2022,Birrer_2022}. Time-delay cosmography has been applied to supernovae lensed by clusters~\citep{Grillo_2024, Suyu2025, Pierel2025} and to quasars lensed by galaxies, achieving a precision of around 4.6~\% \citep{Birrer_2025}. Galaxy-scale double-plane lensing, on the other hand, refers to the alignment of three galaxies, such that we observe light from an intermediate redshift galaxy (plane 2) lensed by a foreground galaxy (plane 1), and light from an even more distant third galaxy (plane 3) lensed by both of the others. This enables a measurement of the so-called \emph{cosmological scaling factor} $\eta = D_{12}D_3/D_{13}D_2$, where $D_{ml}$ is the angular diameter distance from planes $m$ to $l$. $\eta$ is independent of the Hubble constant, but very sensitive to the expansion history of the universe, and is particularly promising for placing independent constraints on the dark-energy equation of state, $w(z)$, \citep{Collett_2012,Collett_2014,Smith_2022,Sahu_2025,Bowden_2025}. The discovery of the Einstein zig-zag lens J1721+8842 \citep{Dux_2024}, a rare and spectacular double-plane lens with a sextuply imaged quasar source, has renewed interest in the intersection of these fields, i.e. time-delay cosmography with double-plane lenses. Such a system is extremely sensitive to the mass profiles of both lenses, and could provide a simultaneous measurement of $H_0$ and $w(z)$ \citep{Collett_2015b}. In light of the ``Hubble tension''  \citep[see e.g.][for a recent review]{Di_Valentino_2025} and the recent evidence for dynamical dark energy \citep{Karim_2025}, such constraints have never been more important. 

Reconstructing a 3D distribution of matter from its influence on the 2D images we observe is an unsurprisingly challenging task. Any transformation to the mass profile of the lens which preserves the images we see cannot be constrained from these images alone, even when impacting the source position and morphology, as these are never directly observed \citep{Schneider_2013}. Closely linked to this, our distinction between the background cosmology, lens and perturbers is a fundamentally artificial one -- the path of a light beam is determined only by the geometry of the spacetime through which it passes. Any matter field which is sufficiently diffuse and approximately uniform on the angular scales of the image will change only the overall fluxes and length scales we observe. The presence of such a field is therefore indistinguishable from a change to the background cosmology without strong priors or independent constraints on this cosmology, the position and luminosity of the source, or the physical properties of the lens. This is the so-called \emph{mass-sheet degeneracy} (MSD), the most famous of a broader class of source position transforms, and is perhaps the most significant limitation in strong-lens cosmography \citep{Falco_1985,Schneider_2013,Birrer_2016,Kochanek_2020}. As the community races to study J1721+8842 and other such systems, it is crucial that the double-plane MSD is well-understood and accounted for.

In \cite{McCully_2014,McCully_2017,Fleury_2021b, Johnson_2025}, the impact of matter lying along the line of sight in multi-plane systems is studied. In \cite{Schneider_2014}, the impact of the mass-sheet degeneracy on double-plane time delays \emph{with a fixed cosmological scaling factor $\eta$} is derived, as well as showing that the scaling factor cannot be measured without accounting for this degeneracy. In \cite{Schneider_2019} and Teodori (in preparation), these results are generalised to an arbitrary number of lens planes. In Li et al. (in preparation), these results are exploited to place constraints on the dark matter mass profile and stellar-to-halo mass ratio of the foreground lens galaxy.

In this paper, we build on these previous works to consider the impact of the line of sight and mass-sheet degeneracy on constraints on $H_0$ and $\eta$ from double-plane lenses. We point out that careful modelling choices can fully respect the uncertainty arising from these degeneracies, while simplifying the observations needed to correct for them. In \cref{sec:los_effects}, we summarise the double-plane lensing formalism in the presence of line-of-sight convergence, and show how these convergence terms are implicitly absorbed into lensing quantities, including $\eta$, meaning that it cannot simply be fixed from cosmological priors. In \cref{sec:msd}, we summarise the double-plane mass-sheet transformation when $\eta$ is unknown, derive the expression for relative time delays under this transformation, and show that geometrical relationships between angular diameter distances can be used to simplify the impact of the degeneracy. In \cref{sec:implications}, we discuss the implications for $H_0$ constraints, and summarise in \cref{sec:discussion}.

%%%%%%%%%%%%%%%%%%%%%%%%%%%%
\section{Weak lensing corrections to the double-plane lens equation}
\label{sec:los_effects}
%%%%%%%%%%%%%%%%%%%%%%%%%%%%
%
\subsection{Multi-plane lensing}
Following \cite{Schneider_2014}, we consider a distribution of $N$ lenses and sources with redshifts $z_l$. We denote with $D_l$ the angular diameter distance corresponding to $z_l$, and $D_{ml}$ the angular diameter distance to $z_l$ as seen from $z_m$. $D_l$ and $D_{ml}$ refer specifically to distances in the \emph{background homogeneous--isotropic cosmology}, without any additional lensing effects. The projected surface mass density in the $l$\textsuperscript{th} plane $\Sigma_l(\bm{x}_l)$ results in a deflection angle 
\begin{equation}
    \hat{\bm{\alpha}}_l(\bx_l) = \frac{4G}{c^2}\int\rd^2\bx' \,\Sigma_l\pa{\bx'}\frac{\bx_l-\bx'}{|\bx_l-\bx'|^2} \, ,
\end{equation}
where $\bx_l$ is the transverse separation vector in the $l$\textsuperscript{th} plane.

The propagation of a light ray through these $N-1$ planes is determined by both the deflections of the ray by these lenses, whose strengths vary with the corresponding impact parameters $\bx_l$, as well as weak-lensing corrections -- integrated effects arising from matter which is sufficiently diffuse or far from the line of sight to produce only tidal distortions to the ray. In the presence of these effects, the unlensed angular position $\btheta_l$ at which the ray intersects plane $l$ is given by \citep{Fleury_2021b}
\begin{equation}
    \btheta_l = \btheta - \sum^{l-1}_{m=1}\frac{D_{ml}}{D_{l}}\bA^{-1}_{\mathrm{o}l}\bA_{ml}\hat{\bm{\alpha}}_m\pa{\bx_m}. \label{eq:recursion1}
\end{equation}
The distortion matrices $\bA_{ml}$ encode the tidal corrections from matter between planes $m$ and $l$, and are typically decomposed as 
\begin{equation}
    \bA_{ml} = \begin{bmatrix}
  1 - \kappa_{ml} - \mathrm{Re}\left(\gamma_{ml}\right) & -\mathrm{Im}\left(\gamma_{ml}\right) \\
  -\mathrm{Im}\left(\gamma_{ml}\right) & 1 - \kappa_{ml} + \mathrm{Re}\left(\gamma_{ml}\right)
  \end{bmatrix}, \label{eq:Amatrix}
\end{equation}
The convergence $\kappa$ isotropically rescales an image, while the shear $\gamma=\gamma_1+\ri\gamma_2$ distorts the image shape.
\subsection{The double-plane lens equation}
In the following, we consider specifically the ``double-plane'' case where $N=3$, i.e. the alignment of a first lens at $z_1$, a source at $z_3$, and an intermediate-redshift lens at $z_2$ whose light is also distorted by the first lens. We include the effects of line-of-sight convergence terms, but neglect the shear. The lens equations governing such a system can be written as
\begin{gather}
    \btheta_2 = \btheta - \frac{1-\kappa_{12}}{1-\kappa_2}\bar{\bm{\alpha}}_1[(1-\kappa_1)\btheta],
    \label{eq:los1} \\
    \btheta_3 = \btheta - \frac{1-\kappa_{13}}{1-\kappa_3}\bar{\eta}\bar{\bm{\alpha}}_1[(1-\kappa_1)\btheta]-\frac{1-\kappa_{23}}{1-\kappa_3}\bar{\bm{\alpha}}_2[(1-\kappa_2)\btheta_2],
    \label{eq:los2}    
\end{gather}
where $\bar{\eta}$ is the background cosmological scaling factor, given by
\begin{equation}
    \bar{\eta} = \frac{D_2D_{13}}{D_{12}D_3}, \label{eq:bareta}
\end{equation}
and the displacement angles $\bar{\bm{\alpha}}_1(\btheta)$ and $\bar{\bm{\alpha}}_2(\btheta_2)$ are defined as
\begin{gather}
    \bar{\bm{\alpha}}_1(\btheta) = \frac{\rd \bar{\psi}_1(\btheta)}{\rd\btheta} = \frac{D_{12}}{D_2}\uvect{\alpha}_1(D_1\btheta), \label{eq:alpha1_convention} \\
    \bar{\bm{\alpha}}_2(\btheta_2) = \frac{\rd \bar{\psi}_2(\btheta_2)}{\rd\btheta_2} = \frac{D_{23}}{D_3}\uvect{\alpha}_2(D_2\btheta_2), \label{eq:alpha2_convention}
\end{gather}
where $\bar{\psi}_1\pa{\btheta}$ and $\bar{\psi}_2\pa{\btheta_2}$ are the lensing potentials. 
\subsection{Absorbing line-of-sight effects}
The convergence terms appearing in \cref{eq:los1,eq:los2} are unobservable from imaging constraints, and a double-plane system can be modelled without explicitly accounting for them. With this choice, the cosmological scaling factor which we measure is related to the background quantity $\bar{\eta}$ via
\begin{equation}
    \eta \equiv \frac{(1-\kappa_2)(1-\kappa_{13})}{(1-\kappa_{12})(1-\kappa_3)} \, \bar{\eta}, \label{eq:los_eta}
\end{equation}
and the potentials we measure likewise absorb corresponding line-of-sight contributions
\begin{gather}
    \psi_1(\btheta) \equiv \frac{1-\kappa_{12}}{(1-\kappa_1)(1-\kappa_2)}\bar{\psi}_1[(1-\kappa_1)\btheta], \label{eq:lospot_1}\\ 
    \psi_2(\btheta) \equiv \frac{1-\kappa_{23}}{(1-\kappa_2)(1-\kappa_3)}\bar{\psi}_2[(1-\kappa_2)\btheta], \label{eq:lospot_2}
\end{gather}
with corresponding displacement angles
\begin{gather}
    \bm{\alpha}_1(\btheta) \equiv \frac{\rd \psi_1(\btheta)}{\rd\btheta} = \frac{1-\kappa_{12}}{1-\kappa_2}\bar{\bm{\alpha}}_1[(1-\kappa_1)\btheta], \label{eq:alpha1_los} \\ 
    \bm{\alpha}_2(\btheta) \equiv \frac{\rd \psi_2(\btheta_2)}{\rd\btheta_2} = \frac{1-\kappa_{23}}{1-\kappa_3}\bar{\bm{\alpha}}_2[(1-\kappa_2)\btheta]. \label{eq:alpha2_los}
\end{gather}
With these definitions, we can write
\begin{gather}
    \btheta_2 = \btheta - \bm{\alpha}_{1}(\btheta), \label{eq:eff_lens_1} \\
    \btheta_3 = \btheta - \eta\bm{\alpha}_1(\btheta)-\bm{\alpha}_2(\btheta_2), \label{eq:eff_lens_2}
\end{gather}
which are identical in form to the equations governing a double-plane lens system with no line-of-sight effects.
\subsection{Time delays}
\label{sec:los_on_time_delays}
From e.g. \cite{Fleury_2021b}, the light travel time from $\btheta_3$ via $\btheta_2$ and $\btheta$ to the observer is given by 
\begin{equation}
    T(\btheta,\btheta_2,\btheta_3) = \tau_{12} \, \phi_{12}\pa{\btheta,\btheta_2} + \tau_{23} \, \phi_{23}\pa{\btheta_2,\btheta_3}, \label{eq:time_delays}
\end{equation}
where $\phi_{12}\pa{\btheta,\btheta_2}$ and $\phi_{23}\pa{\btheta_2,\btheta_3}$ are Fermat potentials, dimensionless quantities given by
\begin{align}
    \phi_{12}(\btheta,\btheta_2)
    =& \frac{1}{2}(\btheta-\btheta_2)^2-\psi_1\pa{\btheta},\label{eq:phi_1} \\
    \phi_{23}(\btheta_2,\btheta_3) 
    =& \frac{1}{2}(\btheta_2-\btheta_3)^2-\psi_2\pa{\btheta_2}; \label{eq:phi_2}
\end{align}
while $\tau_{12}$ and $\tau_{23}$ are the time-delay scales, a combination of background angular-diameter distances and convergence factors,
\begin{gather}
    \tau_{12} = \frac{1}{c}(1+z_1)\frac{(1-\kappa_1)(1-\kappa_2)}{1-\kappa_{12}}\frac{D_1D_2}{D_{12}} = \pa{1-\kappa\h{LOS}_{12}}\bar{\tau}_{12}, \label{eq:tau12_los} \\
    \tau_{23} = \frac{1}{c}(1+z_2)\frac{(1-\kappa_2)(1-\kappa_3)}{1-\kappa_{23}}\frac{D_2D_3}{D_{23}} = \pa{1-\kappa\h{LOS}_{23}}\bar{\tau}_{23}, \label{eq:tau23_los}
\end{gather}
with
\begin{equation}
    \bar{\tau}_{ml} \equiv \frac{1}{c}(1+z_m)\frac{D_mD_l}{D_{ml}}, \qquad 1-\kappa\h{LOS}_{ml} \equiv \frac{(1-\kappa_m)(1-\kappa_l)}{1-\kappa_{ml}}. \label{eq:tau_ml}
\end{equation}
Note that, in the literature, $\tau_{ml}$ (or $D^{\Delta t}_{ml}=c \, \tau_{ml}$, the time-delay distance) is often used to refer specifically to the background quantity. For book-keeping purposes and following our previous convention, we use $\bar{\tau}_{ml}$ for the background quantity and $\tau_{ml}$ for the one specific to the line of sight in question. 

%
%%%%%%%%%%%%%%%%%%%%%%%%%%%%
\section{The mass-sheet degeneracy}
\label{sec:msd}
%%%%%%%%%%%%%%%%%%%%%%%%%%%%
%
% The mass-sheet degeneracy is an example of a source-position transform \cite{Schneider_2013} \DPJ{check citations}, whereby our lack of knowledge of the unlensed source position means that this position and the lens potential can be simultaneously rescaled and transformed without impacting the observed image positions. Relative time delays are rescaled, and thus a restrictive lens model choice which artificially breaks this degeneracy will systematically bias values of $H_0$ inferred from time-delay cosmography.

In \cite{Schneider_2014,Schneider_2019}, it is shown that the mass-sheet degeneracy generalises to an arbitrary number of main lens planes. In \cite{Schneider_2014}, which focuses specifically on the double-plane MSD, the effect of the degeneracy on time-delay cosmography with a fixed $\beta$ ($\eta^{-1}$ in the conventions of this paper), and on the inference of $\beta$ from compound lenses, are considered. In this section, we explore the implications of a non-fixed $\eta$ for time-delay cosmography in light of the MSD.
\subsection{A recap of the double-plane mass-sheet degeneracy}
\label{sec:msd_lens_observables}
In the convention from \cref{sec:los_effects} that lensing quantities are scaled relative to the subsequent plane, the mass-sheet-transformed first lens displacement angle takes the familiar form
\begin{equation}
    \tilde{\bm{\alpha}}_1(\btheta) = \lambda\bm{\alpha}_1(\btheta)+\pa{1-\lambda}\btheta, \label{eq:full_alpha_MSD_1}
\end{equation}
which satisfies \cref{eq:eff_lens_1} up to a rescaling by $\lambda$ of $\bm{\theta}_2$ . In other words, if a displacement profile $\bm{\alpha}_1(\btheta)$ satisfies \cref{eq:eff_lens_1}, then so too will $\tilde{\bm{\alpha}}_1(\btheta)$. While $\bm{\theta}_2$ in such a model would be inferred as $\tilde{\bm{\theta}}_2=\lambda\bm{\theta}_2$, this change would not be observable from the image positions we have access to.

\cite{Schneider_2014} asks and answers the question of whether there are transformations $\bm{\alpha}_2(\btheta_2) \rightarrow \tilde{\bm{\alpha}}_2(\tilde{\btheta}_2)$ and $\eta \rightarrow \tilde{\eta}$ which satisfy a transformed version of \cref{eq:eff_lens_2},
\begin{equation}
    \nu_3\btheta_3 = \btheta - \tilde{\eta}\tilde{\bm{\alpha}}_1(\btheta) - \tilde{\bm{\alpha}}_2(\tilde{\btheta}_2),
\end{equation}
i.e. transformations that preserve the (unknown) source position up to a uniform rescaling by $\nu_3$, and in which $\tilde{\bm{\alpha}}_1(\btheta)$ must also satisfy \cref{eq:eff_lens_1}, and thus takes the form~\eqref{eq:full_alpha_MSD_1}. This condition is satisfied by the transformed quantities
\begin{equation}
    \nu_3 = \frac{\tilde{\eta}-1}{\eta-1}, \label{eq:nu_3} 
\end{equation}
and
\begin{equation}
    \tilde{\bm{\alpha}}_2(\tilde{\btheta}_2)  = \nu_3\bm{\alpha}_2\pa{\lambda^{-1}\tilde{\btheta}_2}+\tilde{\eta}\pa{1-r_\eta\nu_3\lambda^{-1}}\tilde{\btheta}_2, \label{eq:full_alpha_MSD_2}
\end{equation}
where we have introduced $r_\eta \equiv \eta/\tilde{\eta}$. The lensing potentials of the two lenses under this transformation are
\begin{align}
    \tilde{\psi}_1(\btheta) &= \lambda\psi_1(\btheta)+\frac{1}{2}\pa{1-\lambda}\btheta^2, \label{eq:full_psi_MSD_1}\\
    \tilde{\psi}_2(\tilde{\btheta}_2) &= \lambda \nu_3\psi_2\pa{\lambda^{-1}\tilde{\btheta}_2}+\frac{1}{2}\tilde{\eta}\pa{1-r_\eta\nu_3\lambda^{-1}}\tilde{\btheta}^2_2. \label{eq:full_psi_MSD_2}
\end{align}
$\tilde{\eta}$, not to be confused with the background quantity $\bar{\eta}$ in \cref{eq:bareta}, should be understood as the scaling factor needed to satisfy the lens equations under a mass-sheet transformation of the two lens potentials; it is either fixed or left free within the lens modelling, and need not correspond to the physical value of $\eta$.
\subsection{Transformed Fermat potentials with a flexible scaling factor}

The Fermat potential that would be predicted from image positions under a mass-sheet transformation for the first lens plane is identical to that known in single-plane lensing, and takes the form
\begin{align}
    \tilde{\phi}_{12}(\btheta,\tilde{\btheta}_2)
    &= \frac{1}{2}(\btheta-\tilde{\btheta}_2)^2-\tilde{\psi}_1\pa{\btheta}, \\
    &= \lambda\phi_{12}(\btheta,\btheta_2)+\frac{1}{2}\lambda\pa{\lambda-1}\btheta_2^2. \label{eq:msd_phi_1}
\end{align}
For the second lens plane, we would have
\begin{align}
    \tilde{\phi}_{23}(\tilde{\btheta}_2,\tilde{\btheta}_3) 
    &= \frac{1}{2}(\tilde{\btheta}_2-\tilde{\btheta}_3)^2-\tilde{\psi}_2\pa{\tilde{\btheta}_2}, \\ 
    &= \lambda\nu_3\phi_{23}(\btheta_2,\btheta_3) + \frac{1}{2}\nu_3\pa{\nu_3-\lambda}\btheta^2_3 \notag \\
    & \quad +\frac{1}{2}\lambda\pac{\lambda\pa{1-\tilde{\eta}}-\nu_3\pa{1-r_\eta\tilde{\eta}}}\btheta^2_2 , \label{eq:msd_phi_2}
\end{align}
where $\phi_{12}(\btheta,\btheta_2)$ and $\phi_{23}(\btheta_2,\btheta_3)$ are the untransformed quantities defined in \cref{eq:phi_1,eq:phi_2}. In the case where $\eta$ is correctly fixed (i.e. $\tilde{\eta}=\eta$), \cref{eq:msd_phi_2} reduces to eq. (16) in \cite{Schneider_2014}.

\subsection{The transformed time-delay distance}

Putting it all together, and ignoring the term in \cref{eq:phi_2} which depends only on $\btheta_3$, and which is therefore unobservable from relative time delays between different image positions, the overall time delay function reads
\begin{align}
    \tilde{T}(\btheta, \tilde{\btheta}_2, \tilde{\btheta}_3) &= \ \tilde{\tau}_{12}\tilde{\phi}_{12}(\btheta,\tilde{\btheta}_2)+\tilde{\tau}_{23}\tilde{\phi}_{23}(\tilde{\btheta}_2,\tilde{\btheta}_3), \notag \\
    &= \ \tilde{\tau}_{12}\lambda\phi_{12}(\btheta,\btheta_2)+\tilde{\tau}_{23}\lambda\nu_3\phi_{23}(\btheta_2,\btheta_3)
    \notag \\
    & \quad +\frac{1}{2}\tilde{\tau}_{12}\lambda\pa{\lambda-1}\btheta_2^2 
    \notag \\
    & \qquad +\frac{1}{2}\tilde{\tau}_{23}\lambda\pac{\lambda\pa{1-\tilde{\eta}}-\nu_3\pa{1-r_\eta\tilde{\eta}}}\btheta^2_2, \label{eq:time_delay_unsubstituted}
\end{align}
where $\tilde{\tau}_{ml}$ refers to the time-delay scales we would (mis-)infer from (i) the actual, measured time delays and (ii) the mass-sheet transformed Fermat potentials~$\tilde{\phi}$, themselves (mis-)inferred from imaging data with an overly-restrictive mass model. 

\subsection{Simplifications with the unfolding relation}
The time-delay scales in \cref{eq:tau12_los,eq:tau23_los} are related via the so-called ``unfolding relation'',
\begin{gather}
    \tau^{-1}_{13} = \tau^{-1}_{12}+\tau^{-1}_{23}, \label{eq:unfolding_relation}
\end{gather}
which crucially is true not only in a homogeneous--isotropic cosmological background \citep{Schneider_1992}, but in any background spacetime, and hence also in the presence of weak-lensing corrections \citep{Fleury_2021b}. Furthermore, from \cref{eq:bareta,eq:los_eta,eq:tau12_los,eq:tau23_los} we have that $\eta = \tau_{12}/\tau_{13}$, and so
\begin{equation}
    \tau_{12} = \pa{\eta-1}\tau_{23}. \label{eq:tau23_to_tau12} 
\end{equation}
Using this relationship, we have the freedom to re-express the time-delay equation in terms of $\tau_{12}$ or $\tau_{23}$ only,
\begin{align}
    T(\btheta, \btheta_2, \btheta_3)
    &= \tau_{12}\pa{\phi_{12}+\frac{1}{\eta-1}\phi_{23}}, \label{eq:tau12_factored} \\
    &= \tau_{23}\pac{\pa{\eta-1}\phi_{12}+\phi_{23}}.
    \label{eq:tau23_factored}
\end{align}
Now, while each of the terms appearing in \cref{eq:tau23_to_tau12} may be mis-measured under a mass-sheet degeneracy, a self-consistent model must enforce this relationship between them, and so we must also have
\begin{equation}
    \tilde{\tau}_{12} = \pa{\tilde{\eta}-1}\tilde{\tau}_{23}. \label{eq:tau23_to_tau12_MSD} 
\end{equation}
Using \cref{eq:tau23_to_tau12_MSD}, just as in the case where no line-of-sight corrections are present and $\eta$ is fixed \citep{Schneider_2014}, the terms in \cref{eq:time_delay_unsubstituted} proportional to $\btheta_2$ cancel out fully. Replacing $\nu_3$ with its expression in \cref{eq:nu_3}, we are left with
\begin{equation}
    \tilde{T}(\btheta, \tilde{\btheta}_2, \tilde{\btheta}_3)
    = \lambda\pac{\tilde{\tau}_{12} \phi_{12}(\btheta,\btheta_2)+\pa{\frac{\tilde{\eta}-1}{\eta-1}}\tilde{\tau}_{23}\phi_{23}(\btheta_2,\btheta_3)}. \label{eq:time_delay_unsubstituted_MSD}
\end{equation}
If, as before, we choose to express the entire time-delay function in terms of a single time-delay scale, we may write
\begin{empheq}[box=\fbox]{align}
    \tilde{T}(\btheta, \tilde{\btheta}_2, \tilde{\btheta}_3)
    &= \lambda\tilde{\tau}_{12}\pac{\phi_{12}(\btheta,\btheta_2)+\pa{\frac{1}{\eta-1}}\phi_{23}(\btheta_2,\btheta_3)} \\ 
    &= \lambda\pa{\frac{\tilde{\tau}_{12}}{\tau_{12}}} \, T(\btheta,\btheta_2,\btheta_3),  \label{eq:tau12_factored_msd} 
\end{empheq}
or, if we prefer to factor out $\tau_{23}$,
\begin{align}
    \tilde{T}(\btheta, \tilde{\btheta}_2, \tilde{\btheta}_3)  
    &= \lambda\tilde{\tau}_{23}\pac{\pa{\tilde{\eta}-1}\phi_{12}(\btheta,\btheta_2)+\pa{\frac{\tilde{\eta}-1}{\eta-1}}\phi_{23}(\btheta_2,\btheta_3)}, \\ 
    &= \lambda\pa{\frac{\tilde{\tau}_{23}}{\tau_{23}}}\pa{\frac{\tilde{\eta}-1}{\eta-1}} \, T(\btheta,\btheta_2,\btheta_3). \label{eq:tau23_factored_msd} 
\end{align}
It is then clear that the expressions in \cref{eq:time_delay_unsubstituted_MSD,eq:tau12_factored_msd,eq:tau23_factored_msd} are proportional to the one in \cref{eq:time_delay_unsubstituted}. Put differently, the cancellation of the terms involving $\bm{\theta}_2$ in \cref{eq:time_delay_unsubstituted} ensures that time-delay ratios are conserved under a mass-sheet transformation.
%
%%%%%%%%%%%%%%%%%%%%%%%%%%%%
\section{Implications for time-delay cosmography}
\label{sec:implications}
%%%%%%%%%%%%%%%%%%%%%%%%%%%%
%
Angular-diameter distances, when expressed in terms of redshifts, are inversely proportional to the Hubble constant, and thus, from \cref{eq:tau_ml,eq:time_delay_unsubstituted}, so too is the time-delay function. With relative time-delay observations and Fermat potentials constrained from imaging, time-delay cosmography can therefore provide constraints on $H_0$. However, this measurement will be degenerate with any factor which rescales relative time delays and which cannot be observed from lens images, unless the factor can be independently constrained.

Now, the implications of \cref{eq:tau23_to_tau12_MSD} are significant: even when subject to line-of-sight effects or a classic mass-sheet transformation, constraints on any two of $\tilde{\tau}_{12}$, $\tilde{\eta}$ and $\tilde{\tau}_{23}$ fully determines the third. In practice, the consequences of this are evident in the re-expression of \cref{eq:time_delay_unsubstituted_MSD} as \cref{eq:tau12_factored}. A measurement of $\tilde{\eta}$ from imaging observables, even with no cosmological priors or knowledge of the mass-sheet transformation or line-of-sight terms at play\footnote{Note again that $\tilde{\eta}$ is a quantity within the lens model, which can generally be constrained to high precision \citep{Collett_2014,Sahu_2025,Bowden_2025}.}, reduces the uncertainty in the Hubble constant to the uncertainty in $\lambda$, $\kappa\h{LOS}_{12}$, and the measurement error in $\tilde{T}(\btheta,\tilde{\btheta}_2,\tilde{\btheta}_3)$, i.e.\footnote{The reader might be more familiar with the relation $H_0/H_0\h{mes}~\propto \lambda$. This is purely a matter of convention, whether $\lambda$ is understood either as transforming true quantities into their mis-measured counterparts (as in this article) or vice-versa (as is often the case in the single-plane lensing literature).}
\begin{equation}
    H_0 = \lambda^{-1}\pa{1-\kappa\h{LOS}_{12}}H_0\h{mes}.
\end{equation}
$\lambda$ can be constrained from the dynamical modelling of the foreground lens galaxy with resolved velocity dispersion measurements, and $\kappa\h{LOS}_{12}\approx 1-\kappa_1-\kappa_2+\kappa_{12}$ can be estimated from this modelling and from galaxy number counts up to $z_2$ \citep{Rusu_2019} or weak lensing \citep{Tihhonova_2018, Tihhonova_2020}. Provided we impose the relationship in \cref{eq:tau23_to_tau12_MSD}, we do not need any knowledge of the true value of $\eta$, nor $\kappa\h{LOS}_{23}$, and hence weak-lensing corrections from objects beyond $z_2$ do not contribute to our uncertainty in $H_0$, even when the source is located at $z_3$. A mass-sheet transformation of $\eta$ and our ignorance of  $\kappa\h{LOS}_{23}$ will result in other aspects of the lens model being misinferred, and limit our knowledge of other cosmological parameters, but these errors will not propagate into the constraints on $H_0$.
%
%%%%%%%%%%%%%%%%%%%%%%%%%%%%
\section{Discussion}
\label{sec:discussion}
%%%%%%%%%%%%%%%%%%%%%%%%%%%%
%
In this paper, we have discussed the implications of the double-plane mass-sheet degeneracy for time-delay cosmography. In \cref{sec:los_effects}, we summarised the equations governing double-plane lensing in the presence of tidal weak lensing convergence corrections from the line of sight. We argued that each of the quantities which appear in the lens and time-delay equations, namely the lensing potentials $\psi_{ml}$ and their derivatives, the time-delay scales $\tau_{ml}$ and the scaling factor $\eta$, should be understood as a combination of a background quantity and a convergence correction. While these corrections are easily absorbed into the lensing potentials, there is no reason that the scaling factor and time-delay scales should respect a simple internally-consistent cosmological model without an allowance for a line-of-sight correction. 

In \cref{sec:msd}, we discussed the impact of the mass-sheet degeneracy on double-plane lensing. Building on the results of \cite{Schneider_2014}, we derived an expression for the impact of this degeneracy on measured time delays when $\eta$ is not fixed, but instead measured from lensing observables. We pointed out that, regardless of the line-of-sight convergence terms present, the time-delay scales can be related via the cosmological scaling factor, which means that we can re-express the time-delay function in terms of a single time-delay scale of our choice. 

In \cref{sec:implications}, we considered the implications of these results for time-delay cosmography. Given the relationship between the time-delay scales and $\eta$, we argued that the precision of an $H_0$ measurement is limited only by the uncertainties in relative time delays and lens model constraints from imaging (with an artificially fixed mass sheet), the mass-sheet transformation $\lambda$ in the first plane, and the weighted line-of-sight contribution $\kappa\h{LOS}_{12}$ from matter located between the observer and the redshift of the first source/second lens. We do not need to know the true value of the cosmological scaling factor~$\eta$, nor the line-of-sight contribution between the first and second source planes, nor the true mass profile of the second plane.

In practice, it is not usually $\eta$ and a particular $\tau_{ml}$ which is parameterised, but instead parameters such as $H_0$ and $\Omega\e{m}$ of a cosmological model, with a single external convergence term constrained from galaxy number counts and/or velocity dispersion measurements. The relationships we have presented and discussed simply serve to illustrate the two main messages of this paper. The first is that, in the case of double-plane lensing, a background cosmological model for $\eta$ and $\tau_{ml}$ and a single mass-sheet correction term is insufficient, as there is a distinct and unpredictable line-of-sight contribution to each of these terms. Nonetheless, they remain mutually dependent, and our second message is that the relationship between the time-delay scales and $\eta$ should be exploited to safely reduce the freedom of the model. In doing so, the time-delay function can be reduced back to a combination of direct lensing observables and single corrections for the mass-sheet transformation and the line of sight.

This result, while worth illustrating explicitly, is not surprising. The central message of \cite{Schneider_2014} is that the double-plane mass-sheet degeneracy, as in the single-plane case, can be recast as a transformation of the background cosmology within which the system is embedded. It is therefore natural that systematic biases in $\eta$, in which there is no $H_0$ dependence, should not propagate into the constraints on $H_0$, provided the modelled parameters still fit the data. Furthermore, given the result in \cite{Schneider_2014} that the mass-sheet parameter needed to correct the $H_0$ measurement is also that which rescales the lensing potential of the foreground lens, it is inevitable that the relevant ``external convergence'' term is $\kappa\h{LOS}_{12}$. There are two distinct but equivalent ways to understand the contribution of line-of-sight convergence. In the approach adopted in this paper, these convergence terms couple to background angular diameter distances, and all lensing quantities can be understood as being defined in the ``local cosmology'' along that line of sight. As such, cosmological measurements from lensing are measurements of these ``line-of-sight-specific'' cosmologies. Alternatively, line-of-sight convergence terms can be imagined in the same way as the main lens(es), appearing as distinct terms within the lens equation. Given the mass-sheet transformation, we accept that these terms cannot be constrained, and must implicitly re-insert their effect via independent constraints. In either description, their effect on lensing observables is exactly equivalent to an internal mass-sheet transformation. As $\kappa\h{LOS}_{12}$ is the term which serves this role in the first plane (see eg. Fleury et al. \citeyear{Fleury_2021a}), following the result of \cite{Schneider_2014}, it is inevitable that it should be the term to accompany $\lambda$ in limiting our constraints on $H_0$. 

Despite this, there is still value in distinguishing between internal and external mass-sheet transformations, as they are not constrained in the same manner. The strength of line-of-sight convergence can be estimated from comparisons of galaxy number counts around the lens system to simulations \citep{Rusu_2017,Suyu_2010}, while constraints on the internal MSD can be made if there are well-motivated priors on the galaxy mass profile. Both can be constrained with velocity dispersion measurements of the lens' stellar population. The result of this work is good news for the constraining power of double-plane systems -- the line-of-sight convergence with respect to lower redshift sources is generally smaller \citep{Bernardeau_1997}, and nearby galaxies are easier to observe than further ones. The combination of $\lambda$ and $\kappa\h{LOS}_{12}$ needed to correct $H_0$ is precisely that which rescales the foreground lens, and can therefore be estimated from velocity dispersion measurements of this lens alone. Velocity dispersion measurements of the second lens plane could add further precision, as well as providing a semi-independent check on those constraints.

% One may naturally ask, why bother with rare double-plane systems for time-delay cosmography? We have shown that the mass-sheet degeneracy reduces to the same combination of terms acting in single plane lensing. However, our measurement is also limited by our modelling constraints on the mass profiles of both lenses, and that of the  background lens is generally much more weakly constrained than that of a single plane lens. The advantage of double-plane systems lies in the specificity of the mass-sheet transformation. In the single-plane case, a very wide range of reasonable mass profiles could mimic the transformation (citations). On the other hand, the stronger the mass-sheet transformation in the double-plane case, the more the two lens profiles must differ from one another, as $\lambda$ has an inverse effect in both planes \cite{Schneider_2014,Schneider_2019}. This fact can be exploited to limit the impact of the degeneracy, and is the subject of (cite our other paper in preparation).
%

%\clearpage

%\newpage 

%%%%%%%%%%%%%%%%%%%%%%%%%%%%%%%%%%%%%%%%%%%%%%%%%%%%%%%%%%%%%%
\begin{acknowledgements}
The authors thank Tom Collett, Tian Li, Wolfgang Enzi and Giacomo Queirolo for very helpful discussions, and Luca Teodori for comments on an earlier draft of the article. DJ acknowledges support by the First Rand Foundation, South Africa, and the Centre National de la Recherche Scientifique of France.
PF acknowledges support from the French \emph{Agence Nationale de la Recherche} through the ELROND project (ANR-23-CE31-0002). MM acknowledges support by the SNSF (Swiss National Science Foundation) through return CH grant P5R5PT\_225598 and Ambizione grant PZ00P2\_223738. 
\end{acknowledgements}

\section*{\href{https://www.elsevier.com/authors/policies-and-guidelines/credit-author-statement}{CRedIT} authorship contribution statement}

\noindent \textbf{Daniel Johnson:} Conceptualisation, Methodology, Formal Analysis, Writing - Original draft. 
\textbf{Pierre Fleury:} Methodology, Validation, Formal Analysis, Writing -- Review \& Editing.
\textbf{Martin Millon:} Conceptualisation, Methodology, Writing - Review \& Editing

%%%%%%%%%%%%%%%%%%%%%%%%%%%%
% \bibliographystyle{aa.bst}
\bibliographystyle{aa}
\bibliography{main.bib}
\begin{appendix}
%%%%%%%%%%%%%%%%%%%%%%%%%%%%%%%%%%%%%%%%%%%%%%%%%%%%%%%%%%%%%%%
% In the PDF output, floats should be placed
% under their own appendix, not before the title, nor after the
% title of the next appendix.

% In short appendices, onecolumn floats (\figure*
% or \table*) will generate a blank page.
% To prevent this behaviour, a few examples are provided here. 

% In case you have a lot of floating objects for little text and the 
% LaTeX engine moves the floats away from their context, the command
% \FloatBarrier of the “placeins” package will empty the
% float buffer and place all stored floats in the continuity.

% If you still encounter problems with wide floats placement,
% just use the onecolumn environment throughout the appendices.
%%%%%%%%%%%%%%%%%%%%%%%%%%%%%%%%%%%%%%%%%%%%%%%%%%%%%%%%%%%%%%%

%____________________________________________________________
%       Wide floats at the start of an appendix: first method
%-------------------------------------------------------------
% To prevent a blank page after the start of an appendix:
% - Switch to one \onecolumn first
% - Declare the section title
% - Declare the onecolumn float with the parameter [ht!]
% - Revert to \twocolumn at the end of the section
% \onecolumn
\section{Corresponding expressions with the third plane as a reference}
In this article, we have adopted the convention of defining lensing quantities in a given plane with respect to the subsequent plane, meaning that $\bm{\alpha}_1(\btheta)$ is defined with respect to plane 2, and $\bm{\alpha}_2(\btheta_2)$ with respect to plane 3 - we will refer to this as the $\eta$ convention. An alternative approach which is often adopted is to instead define all quantities with respect to the final source plane - the $\beta$ convention. In this section, we briefly summarise the relevant equations of this article in this convention.
\subsection{Lensing quantities}
In the $\beta$ convention, the lens equations in the presence of line-of-sight effects are
\begin{gather}
    \btheta_2 = \btheta - \beta\bm{\alpha}_{1}(\btheta), \label{eq:eff_lens_1_beta} \\
    \btheta_3 = \btheta - \bm{\alpha}_1(\btheta)-\bm{\alpha}_2(\btheta_2), \label{eq:eff_lens_2_beta}
\end{gather}
where
\begin{gather}
    \bm{\alpha}_1(\btheta) \equiv \frac{\rd \psi_1(\btheta)}{\rd\btheta} = \frac{(1-\kappa_{13})D_{13}}{(1-\kappa_3)D_3}\uvect{\alpha}_1\pac{(1-\kappa_1)D_1\btheta}, \label{eq:alpha1_los_beta} \\ 
    \bm{\alpha}_2(\btheta) \equiv \frac{\rd \psi_2(\btheta_2)}{\rd\btheta_2} = \frac{(1-\kappa_{23})D_{23}}{(1-\kappa_3)D_3}\hat{\bm{\alpha}}_2\pac{(1-\kappa_2)D_2\btheta}, \label{eq:alpha2_los_beta}
\end{gather}
and
\begin{equation}
    \beta = \frac{(1-\kappa_{12})(1-\kappa_3)}{(1-\kappa_2)(1-\kappa_{13})}\frac{D_{12}D_3}{D_2D_{13}}. \label{eq:eta_kappa_beta}
\end{equation}
The time delay in \cref{eq:time_delays} and the associated definitions of the Fermat potentials and time-delay distances remain identical, except that, following its redefinition with respect to the third plane, $\psi_1(\btheta)$ must be rescaled, and so \cref{eq:phi_1} is replaced by
\begin{equation}
    \phi_{12}(\btheta,\btheta_2)
    = \frac{1}{2}(\btheta-\btheta_2)^2-\beta\psi_1\pa{\btheta},\label{eq:phi_1_beta}
\end{equation}
\subsection{The mass-sheet degeneracy}
The mass-sheet transformation which preserves lensing observables in the $\beta$ convention are presented in \cite{Schneider_2014}. The transformed displacement angles take the form
\begin{gather}
    \tilde{\bm{\alpha}}_1(\btheta) = \lambda\bm{\alpha}_1(\btheta)+\frac{1-r_\beta\lambda}{\tilde{\beta}}\btheta, \label{eq:full_alpha_MSD_1_tilde_beta}\\
    \tilde{\bm{\alpha}}_2(\tilde{\btheta}_2)  = \nu_3\bm{\alpha}_2\pa{r_\beta^{-1}\lambda^{-1}\tilde{\btheta}_2}+\frac{1-\lambda^{-1}\nu_3}{\tilde{\beta}}\tilde{\btheta}, \label{eq:full_alpha_MSD_2_tilde_beta}
\end{gather}
with
\begin{equation}
    r_\beta = \frac{\tilde{\beta}}{\beta},
\end{equation}
and 
\begin{gather}
    \nu_3 = \frac{\beta}{\tilde{\beta}}\frac{1-\tilde{\beta}}{1-\beta}.
\end{gather}
Once again, we can take advantage of the unfolding relation to write
\begin{equation}
    \frac{\tilde{\tau}_{23}}{\tilde{\tau}_{12}} = \frac{\tilde{\beta}}{1-\tilde{\beta}}, \label{eq:tau_23_to_tau12_MSD_beta}
\end{equation}
which gives 
\begin{align}
    \tilde{T}(\btheta,\tilde{\btheta}_2,\tilde{\btheta}_3)
    &= \lambda\pac{\frac{\tilde{\beta}}{\beta}\tilde{\tau}_{12}\phi_{12}(\btheta,\btheta_2)+\frac{1-\tilde{\beta}}{1-\beta}\tilde{\tau}_{23}\phi_{23}(\btheta_2,\btheta_3)}, \label{eq:beta_td_1}\\
    &= \pa{\frac{\tilde{\beta}}{\beta}}\lambda\pa{\frac{\tilde{\tau}_{12}}{\tau_{12}}} T(\btheta,\btheta_2,\btheta_3), \\
    &= \frac{1-\tilde{\beta}}{\beta}\lambda\pa{\frac{\tilde{\tau}_{23}}{\tau_{23}}} T(\btheta,\btheta_2,\btheta_3).
\end{align}
These expressions are a little more awkward than those in the $\eta$ convention. Nonetheless, the two prefactors involving $\beta$ and $\tilde{\beta}$ within \cref{eq:beta_td_1} are linearly independent. Our uncertainty as to the true value of $\beta$ will once again limit our ability to constrain other cosmological parameters, but does not directly impact the $H_0$ constraint.

\end{appendix}
\end{document}